\documentclass[prb,preprint,amsmath,amssymb,superscriptaddress]{revtex4}
\usepackage{float}
\usepackage{amssymb}
\usepackage{amsfonts}
\usepackage{euscript}
\usepackage{hhline}
\usepackage{pslatex}
\usepackage{tabularx}
\usepackage{ulem}
\usepackage{color}

\usepackage{graphicx}
\usepackage{dcolumn}
\usepackage{bm}
\usepackage[sort&compress]{natbib}

\makeatletter
\renewcommand{\@biblabel}[1]{#1. }
\renewcommand{\@dotsep}{500}
\renewcommand{\@pnumwidth}{0em}
\renewcommand{\l@figure}[2]{
\@dottedtocline{1}{1.5em}{2em}{Figure #1}{}\vspace{15pt}}

\begin{document}

\begin{abstract}
\textbf{Enhancing light-matter interactions on a chip is of paramount importance to study nano- and quantum optics effects and for the realisation of integrated devices, for instance, for classical and quantum photonics, sensing and energy harvesting applications. Engineered nano-devices enable the efficient confinement of light and, ultimately, the control of the spontaneous emission dynamics of single emitters, which is crucial for cavity quantum electrodynamics experiments and for the development of classical and quantum light sources. Here, we report on  the demonstration of enhanced light-matter interaction and Purcell effects on a chip, based on bio-inspired aperiodic devices fabricated in silicon nitride and gallium arsenide. Internal light sources, namely optically-active defect centers in silicon nitride and indium arsenide single quantum dots, are used to image and characterize, by means of micro-photoluminescence spectroscopy, the individual optical modes confined by photonic membranes with Vogel-spiral geometry. By studying the statistics of the measured optical resonances, in partnership with rigorous multiple scattering theory, we observe log-normal distributions and report quality factors with values as high as 2201$\pm$443. Building on the strong light confinement achieved in this novel platform, we further investigate the coupling of single semiconductor quantum dots to the confined optical modes. Our results show cavity quantum electrodynamics effects providing strong modifications of the spontaneous emission decay of single optical transitions. In particular, thanks to the significant modification of the density of optical states demonstrated in Vogel-spiral photonic structures, we show control of the decay lifetime of single emitters with a dynamic range reaching 20. Our findings improve the understanding of the fundamental physical properties of light-emitting Vogel-spiral systems, show their application to quantum photonic devices, and form the basis for the further development of classical and quantum active devices that leverage the unique properties of aperiodic Vogel spiral order on a chip.}
\end{abstract}

\title{Control of single quantum emitters in bio-inspired \\aperiodic nano-photonic devices} 

\author{Oliver J. Trojak}
\thanks{These authors contributed equally to this work}
\affiliation{Department of Physics and Astronomy, University of Southampton, Southampton SO17 1BJ, United Kingdom}
\author{Sean Gorsky}
\thanks{These authors contributed equally to this work}
\affiliation{Department of Electrical \& Computer Engineering and Photonics Center, Boston University, 8 Saint Mary's St., Boston, Massachusetts, 02215, USA.}
\author{Connor Murray}
\affiliation{Department of Physics and Astronomy, University of Southampton, Southampton SO17 1BJ, United Kingdom}
\author{Fabrizio Sgrignuoli}
\affiliation{Department of Electrical \& Computer Engineering and Photonics Center, Boston University, 8 Saint Mary's St., Boston, Massachusetts, 02215, USA.}
\author{Felipe Arruda Pinheiro}
\affiliation{Instituto de Fisica, Universidade Federal do Rio de Janeiro, Rio de Janeiro-RJ 21941-972, Brazil}

\author{Suk-In Park}
\affiliation{Center for Opto-Electronic Materials and Devices Research, Korea Institute of Science and Technology, Seoul 136-791, South Korea}
\author{Jin Dong Song}
\affiliation{Center for Opto-Electronic Materials and Devices Research, Korea Institute of Science and Technology, Seoul 136-791, South Korea}

\author{Luca Dal Negro}\email{dalnegro@bu.edu}
\affiliation{Department of Electrical \& Computer Engineering and Photonics Center, Boston University, 8 Saint Mary's St., Boston, Massachusetts, 02215, USA.}
\affiliation{Division of Material Science \& Engineering, Boston University, 15 Saint Mary's St. Brookline, Massachusetts, 02446, USA}
\affiliation{Department of Physics, Boston University, 590  Commonwealth Ave., Boston, Massachusetts, 02215, USA}
\author{Luca Sapienza}\email{l.sapienza@soton.ac.uk}
\affiliation{Department of Physics and Astronomy, University of Southampton, Southampton SO17 1BJ, United Kingdom}
\maketitle
\date{\today}
Engineered photonic devices play a key role in the development of on-chip sources of classical and quantum light for a variety of applications including nano-lasers \cite{nanolasers}, single-photon sources \cite{Igor}, sensors \cite{sensors}, and energy harvesters \cite{energy}. For example, photonic crystal cavities have reached light confinement quality factors of the order of several millions \cite{high_Q} and, thanks to the strong enhancement of the light-matter interaction, have allowed to study light-matter hybridisation \cite{strong_coupling}, opto-mechanical effects \cite{optomechanics} and to obtain bright quantum light emission on a chip \cite{bright_SPS, bright_SPS2}. However, given the nanometer-scale accuracy required in the device fabrication, their scalability is limited \cite{PhC}. A different route, compared to highly-engineered devices, makes use of fabrication imperfections as a resource to achieve efficient confinement of light in a scalable device \cite{disorder1, disorder2, Balestri}. While disorder-induced light-localization is relatively easy to achieve in one-dimensional systems \cite{Daozhong, Vollmer, science}, it is more challenging to reach in two dimensions \cite{RiboliOPL,RiboliPRL,Schwartz,Segev}.
Due to the limited deterministic design rules, the applications of disordered systems to optical engineering can also be limited \cite{DalNegro_review}, even though its potential, for instance in sensing \cite{disorder2} and energy harvesting \cite{Vynk}, has been demonstrated.

An alternative approach to achieve light confinement in  devices where light is confined in two and three dimensions in space relies on the concept of aperiodic order \cite{DalNegro_review,Vardeny} that, for instance, characterizes the peculiar geometry of quasi-crystals \cite{Baake1}, i.e. long-range ordered systems that lack translational symmetry. Beyond quasicrystals, aperiodic order can be used to engineer novel photonic devices with well-defined deterministic mathematical rules \cite{Vardeny,DalNegro_elliptic,DalNegro_crystals}, providing compatibility with planar nano-fabrication technologies \cite{DalNegro_Book}, as well as distinctive spectral and optical properties \cite{DalNegro_Fractional,Barthelemy,Gellermann,TrevinoOx}, characteristics that make them appealing for nano-photonics applications \cite{Mahler,Guo,Notomi,Macia}.

Here, we demonstrate the potential of nano-photonic devices characterised by a bio-inspired deterministic aperiodic structure. In particular, we fabricate devices with Vogel-spiral geometry (see Methods section for more details) that were originally introduced in relation to the fascinating geometrical problems of phyllotaxis \cite{DalNegro_crystals}. This particular deterministic aperiodic structure was intensively studied in the last decade, and was found to have interesting fundamental optical properties for nano-photonic and nano-plasmonic applications, including: vector-wave localization \cite{SgrignuoliPRB}, polarization-insensitive light diffraction \cite{TrevinoNano}, enhanced second-harmonic generation \cite{Capretti}, light emission enhancement \cite{Gorsky,Lawrence,Guo,Pecora}, and omni-directional photonic band gaps \cite{Liew,Pollard,Agrawal}. In this work, we show that Vogel spiral active membranes enable to achieve efficient light confinement on a chip and establish a novel aperiodic nano-photonic platform for cavity quantum electrodynamics experiments. 

We experimentally investigate, by means of micro-photoluminescence spectroscopy, the optical properties of suspended silicon nitride (Si$_3$N$_4$) membranes (confining light in the visible range of wavelengths) and of gallium arsenide (GaAs) membranes (confining light in the near-infra-red) where air holes are arranged in a golden-angle Vogel-spiral geometry (see Methods). The devices were designed by considering the local density of optical states (LDOS) computed with an accurate spectral method. The LDOS provides a description of the radiation dynamics of a source embedded into an arbitrary structure, from which the spectral wavelength of high quality factor modes, such as those near the band-gap \cite{TrevinoOx,Liew,SgringuoliACS,John}, can be predicted. Specifically, we utilized the rigorous theory of multi-polar expansion to evaluate the two-dimensional electromagnetic Green tensor 
, from which the LDOS 
is rigorously calculated (for more details, see Supplementary Information).

The LDOS is evaluated for arbitrary arrays of parallel and non-overlapping circular cylinders embedded in a non-absorbing medium. The analysis was limited to the transverse electric polarization only, as this is the polarization for which a band-gap is expected to occur in an air-hole membrane \cite{Joannopoulos}. To take into account  the out-of-plane losses, we have applied an effective refractive index method where the material dispersion of the dielectric material is replaced by the effective index of the fundamental guided mode in the unperturbed (without air-holes) three-dimensional heterostructure \cite{Qiu} (see Supplementary Informations for more details). Fig.\,\ref{Fig1}(a) shows a map of the LDOS as a function of wavelength and air-hole diameters, and (b) depicts one-dimensional cuts of the LDOS map for Si$_3$N$_4$ devices. The corresponding plots for GaAs are illustrated in panels (c-d). Clear photonic band-gap and pseudo-gaps (secondary gaps of smaller amplitudes), identified by bright yellow streaks in Figs.\,\ref{Fig1} (a) and (c), are visible in both configurations and their behaviour is determined by the unique multifractal structural properties of the golden-angle Vogel spiral \cite{TrevinoOx}. For example, there are several peaks, more clearly visible in Figs.\,\ref{Fig1} (b) and (d), which correspond to long-lived modes generated by the first-neighbor distributions of the spiral elements \cite{SgrignuoliPRB}. Localized band-edge modes are formed when ring-shaped regions of similar inter-particle separation $d$ are sandwiched between two regions with different values of $d$, creating a photonic heterostructure \cite{TrevinoOx}. Representative band-edge modes characterized by these features are reported in Fig.\,S2. Moreover, these band-edge modes are often spatially extended, long-lived\cite{SgrignuoliPRB}, and less sensitive to local perturbations. This makes golden-angle Vogel spirals a very appealing photonic platform, due to more robust fabrication tolerances than traditional photonic crystals \cite{Pollard}. The presence of band-gaps despite the relatively low index contrast between silicon nitride and air is related to the long-range order in a nearly-isotropic geometry\cite{Rechtsman,TrevinoOx, Liew,DalNegro_elliptic}. Isotropic gaps also imply reduced group velocity modes and therefore increased light-matter interaction, thus making these devices interesting for non-linear optics applications and for the realisation of low-threshold lasers \cite{Pollard,Joannopoulos}.
The LDOS was evaluated at the center of a spiral composed by 350 air-holes and the optimised design resulted in an average hole separation of 273\,nm for the Si$_3$N$_4$ and and 220\,nm for the GaAs devices. 


We first discuss the properties of the free-standing 340\,nm-thick silicon nitride membranes grown by plasma-enhanced chemical vapour deposition on a silicon substrate, into which 1000 air holes arranged in a golden-angle spiral geometry were etched (see the Methods section, for more details about the growth and the fabrication of the devices).  As previously reported, Si$_3$N$_4$ can emit radiation over a broad range of wavelengths, typically spanning from $\sim$600 to $\sim$850\,nm, once excited with external light sources emitting in the blue or near-ultraviolet \cite{disorder1}. Such intrinsic photo-luminescence can be used as an internal light source to characterize the confined optical modes \cite{disorder1}, sustained by the golden-angle spiral structure. A scanning electron micrograph image of a fabricated device is shown in Fig.\,\ref{Fig2}a. The sample is placed under a confocal optical microscope and excited with a 405\,nm continuous-wave laser with an excitation spot diameter of $\sim$\,2\,$\mu$m, which allows light excitation and collection from specific areas of the nano-photonic devices. The emitted light is collected and sent to a grating spectrometer equipped with a charge-coupled device for spectral characterization (see Fig.\,\ref{Fig2}b).



An example of a collected micro-photoluminescence spectrum is shown in Fig.\,\ref{Fig3}a, where sharp resonances, signature of light confinement, appear above the broad emission from the silicon nitride material. By fitting the resonant peaks with Lorentzian functions, we extract quality factors reaching 2201$\pm$443 (see Fig.\,\ref{Fig3}c), thus exceeding values reported for photonic crystal cavities in silicon nitride operating at visible wavelengths \cite{PhC_vis1, PhC_vis2}. \\
It is important to emphasize that the optical resonances supported by open and planar Vogel spirals are embedded in a three-dimensional environment and they can leak out of the two-dimensional plane, according to their quality factors. As a result light is not fully confined in the spiral plane and the resulting optical resonances are actually three-dimensional electromagnetic quasi-modes \cite{SgrignuoliPRB}.
The aperiodic photonic devices under study provide a large number of optical resonances, resembling the behaviour observed in disordered photonic crystal waveguides, confining light along one dimension in the plane \cite{disorder1}. However, here these resonances are three-dimensional electromagnetic quasi-modes with two-dimensional geometrical support of Vogel spiral arrays, with clear advantages for applications since the higher dimensionality of the present devices provides more easily addressable and larger active areas. It is important to emphasize that the optical resonances supported by open, planar Vogel spirals are embedded in a three dimensional environment so that they can leak out of the system plane. As a result light is not fully confined in the spiral plane and the scattering resonances are actual three-dimensional electromagnetic quasi-modes.\\
We have extensively characterized and tested Si$_3$N$_4$ devices with 165, 170, and 215\,nm diameter of the air holes. Examples of the statistics of the collected optical resonances, plotted as a function of wavelength and quality factor $Q$ (a factor that estimates the quality in the light confinement, evaluated as the ratio between the optical resonance central wavelength and its linewidth), are shown in Fig.\,\ref{Fig4}. We observe that reducing the air-hole diameter (while keeping the rest of the parameters constant) results in a shift of the optical resonances towards longer wavelengths, in agreement with the band-gap calculations shown in Fig.\ref{Fig1}b, and in an increase in the average quality factors (see Supplementary Information for more discussion, including a comparison to simulations). We also show a wide tunability of the wavelength of the confined optical modes over a 100\,nm wavelength window (see Fig.\,\ref{Fig4}a), proving the suitability of aperiodic systems for the realisation of devices with broad-band operation. Concurrently, reducing the air-hole diameters yields an increase in quality factors. This trend is opposite to what is predicted in our simulations, which are only able to predict in-plane quality factor, neglecting the increased out-of-plane losses of more strongly confined modes \cite{boriskina2008}. Thus it is not surprising that higher in-plane (predicted) quality factors would result in lower out-of-plane (measured) quality factors (more discussion on this can be found in the Supplementary material).

Fig.\,\ref{Fig4}b shows that the probability distributions of the measured quality factors follow log-normal statistics. Interestingly, the log-normal distribution of $Q$-factors has been predicted and observed in disordered systems in the Anderson-localized regime \cite{Pinheiro,Weiss,Smolka}. However, differently from traditional random media where the log-normal distribution of $Q$-factor is associated to exponentially localized modes, in Vogel spirals this behaviour is related to the multi-length scale decay of optical resonances, which reflects the multi-fractal complexity of the Vogel-spiral geometry and its LDOS \cite{TrevinoOx} (see Supplementary Informations for more details). 


Now, we use this platform to carry out quantum electro-dynamics experiments by coupling single emitters to the confined optical modes. To this end, we have fabricated golden-angle Vogel-spiral membranes in GaAs, containing a single layer of InAs quantum dots, grown via Stranski-Krastanov technique (for more details about the fabrication process, see the Methods Section). Compared to silicon nitride, GaAs has the advantage of having a higher refractive index (implying a larger contrast with air, thus improving the efficiency in the light confinement via total internal reflection, as shown in Fig.\ref{Fig1}) and can incorporate efficient sources of quantum and classical light to realize novel light-emitting devices, benefiting from the optical properties of aperiodic systems. \\
We show that the emission dynamics of single emitters can be strongly modified due to the modulation of the density of optical states in the aperiodic spiral device (see Supplementary information for more details on LDOS simulations). 

We excite quantum dot emitters, cooled down to cryogenic temperatures, with an above-band 455-nm light-emitting diode and image the confined optical modes on an electron-multiplied charge-coupled device (see Fig.\,2b). In Fig.\,5, examples of photo-luminescence images of the optical modes confined by aperiodic photonic spiral devices are shown: extended annular photonic modes are visible, in accordance with the confined optical modes predicted by the simulations in panel (a) (see Supplementary Informations for more details on the calculations). 

The InAs quantum dots embedded within the GaAs suspended membranes are then excited with a picosecond-pulsed 785\,nm diode laser to carry out time-resolved photo-luminescence spectroscopy. Examples of photo-luminescence spectra are shown in Fig.\,6 (left panels) where sharp peaks, signature of atom-like transitions in single quantum dots are visible. Single emission lines are then spectrally filtered (see Fig.\,2b) and sent to a silicon avalanche photo-diode for photon counting: time-resolved measurements of the emitted intensity are carried out and the experimental results are fitted with exponential decays to extract the lifetimes of the transitions.
When evaluating the decay dynamics of the intensity of the emitted light, we observe strong modifications of the spontaneous emission rate. InAs/GaAs quantum dots located in an unpatterned region of the membrane show lifetimes of $\sim$1\,ns. When coupled to the confined optical modes in our aperiodic photonic devices, instead, we measure lifetimes as short as $\sim$500\,ps: the ratio between the lifetime of quantum dots outside the photonic device and the one measured within the aperiodic structure shows that the spontaneous emission dynamics is Purcell enhanced by a factor 2. We also measure lifetimes of quantum dots within the aperiodic devices as long as 10\,ns, showing a slow-down of the decay dynamics of a factor 10 (see Fig.\,6, right panels). Considering the ratio between the longest and shortest lifetime measured, we show a modification of the spontaneous emission rate of single quantum dots with a dynamic range of 20. Such strong changes in the emission lifetimes reflect the modified density of optical states due to the presence of multifractal  band-gap and pseudo-gaps in bio-inspired Vogel-spiral structures \cite{TrevinoOx}. Our calculations of the predicted Purcell enhancement show that, for optimally positioned emitters \cite{bright_SPS2, Jin, cathodo}, Purcell enhancements up to two-orders of magnitude are theoretically expected, showing the promise of aperiodic devices for strong lifetime control.

In conclusion, our results show that aperiodic nano-photonic devices based on bio-inspired Vogel-spiral geometry are a novel and effective platform for efficient light confinement on a chip and for cavity quantum electrodynamics experiments with single emitters. We foresee the application of such devices for nano-lasers, optical sensors and quantum photonic applications for non-linear optics \cite{DalNegro_review}. In particular, the efficient two-dimensional light confinement and the Purcell enhancement that we have here demonstrated  could be utilized for the development of novel laser and single-photon devices where optical angular momentum is imparted to the emitted radiation \cite{AOM,TrevinoNano,DalNegro_angular}.


\section*{Methods}
\subsection*{Vogel spiral geometry}
Vogel spirals are described by the parametric equations \cite{Naylor}:
\begin{eqnarray}\label{1}
&r_n=a_0\sqrt{n}\\
&\theta_n=n\alpha\label{2}
\end{eqnarray}
where $n=0,1,2,\cdots$ is an integer, $a_0$ is a positive constant called scaling factor, and $\alpha$ is an irrational number, known as the divergence angle, that specifies the constant aperture between successive elements in the spiral \cite{Naylor}. The divergence angle ($\alpha^\circ$ in degree) can be expressed through an irrational number $\xi$ according to the relation $\alpha^\circ=360^\circ-$frac$(\xi)\times360^\circ$, where frac$(\xi)$ denotes the fractional part of $\xi$. Golden-angle Vogel-spiral geometries are generated when $\xi$ is equal to the golden mean $\xi=(1+\sqrt{5})/2$, i.e. when $\alpha\sim137.508^\circ$. 

\subsection*{Silicon Nitride device fabrication}
A 250\,nm-thick silicon nitride layer is deposited on a silicon substrate via plasma-enhanced chemical vapor deposition, using a combination of SiH$_{4}$ and NH$_{3}$ gases at a temperature of 350$^{\circ}$C and a pressure of 750\,mTorr. Electron-beam lithography is then used to write the aperiodic pattern that an inductively coupled plasma reactive ion etch, based on SF$_{6}$ and C$_{4}$F$_{8}$, transfers onto the silicon nitride layer. A KOH wet etch is used to undercut the silicon nitride, creating a free-standing membrane.

\subsection*{Gallium Arsenide device fabrication}
Devices are fabricated on a wafer grown by molecular beam epitaxy, consisting of a single layer of InAs quantum dots grown in the center of a 190\,nm thick layer of GaAs, grown via Stranski-Krastanov technique, on top of a 1\,$\mu$m thick layer of Al$_x$Ga$_{1-x}$As with an average $x$\,=\,0.65, deposited on a GaAs substrate.
Electron-beam lithography is used to write the aperiodic pattern and an inductively coupled plasma reactive ion etch, based on Ar-Cl$_{2}$, is used to transfer it onto the GaAs layer. An HF wet etch is then used to undercut the GaAs, creating a free-standing membrane.

\section*{Acknowledgments}
FAP acknowledges financial support from CNPq, CAPES, and FAPERJ. JDS acknowledges support from IITP grant funded by the Korea government (MSIT No. 20190004340011001). LDN acknowledges partial support from the Army Research Laboratory under Cooperative Agreement Number W911NF-12-2-0023 for the development of theoretical methods utilized in the paper. LS acknowledges partial support from the Royal Society, grant RG170217, the Leverhulme Trust, grant IAF-2019-013, EPSRC, grant EP/P001343/1.

\section*{Author contributions}

LS conceived the optical set-up and built it together with OJT. OJT grew the silicon nitride material, fabricated the silicon nitride and gallium arsenide devices, carried out the experiments and analysed the data, together with LS. CM contributed to the optical characterisation and data analysis of the silicon nitride devices. SIP and JDS grew the quantum dot material. LS, FAP, LDN conceived the research activities and discussed the results, together with the other authors. LS supervised the experimental part of the project and wrote the manuscript with contributions from the other authors.
LDN supervised the design and modeling contributions. SG developed the numerical tools utilized in the paper. SG and FS performed, analyzed, and organized the simulation results. SG wrote the Supplementary Informations with the help of FS and LDN. 

\newpage
\begin{figure}[!h]
\centering
\includegraphics[width=\linewidth]{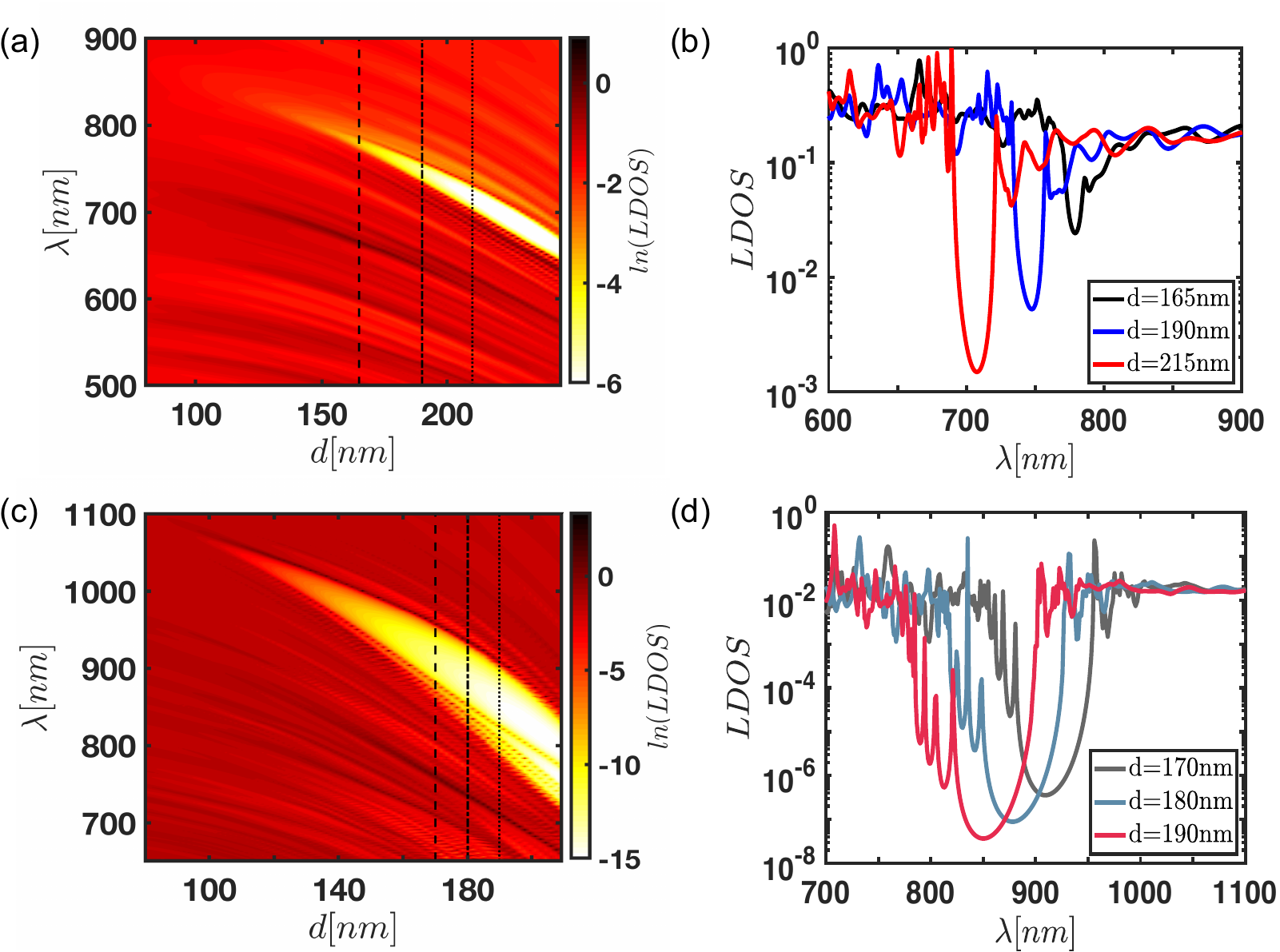}
\caption{Si$_3$N$_4$ and GaAs golden-angle Vogel-spiral device designs. (a) Local density of optical states (LDOS) map calculated at the center of the spiral as a function of the wavelength $\lambda$ and of the air-hole diameters $d$. (b) One dimensional-LDOS line-cuts as a function of wavelength for three different hole diameters: 165 (black line), 190 (blue line), and 215\,nm (red line), for devices with an average inter-particle separation of 273\,nm. Panels (c-d) display the same quantities evaluated for the GaAs devices. Panel (d) shows one dimensional-LDOS line-cut as a function of wavelength for three different air hole diameters: 170  (dark pastel grey line), 180 (dark pastel blue line), and 215\,nm (dark pastel red line), for devices with an average inter-particle separation of 220\,nm.} 
\label{Fig1}
\end{figure}

\newpage
\begin{figure}[h!]
\centering
\includegraphics[width=\linewidth]{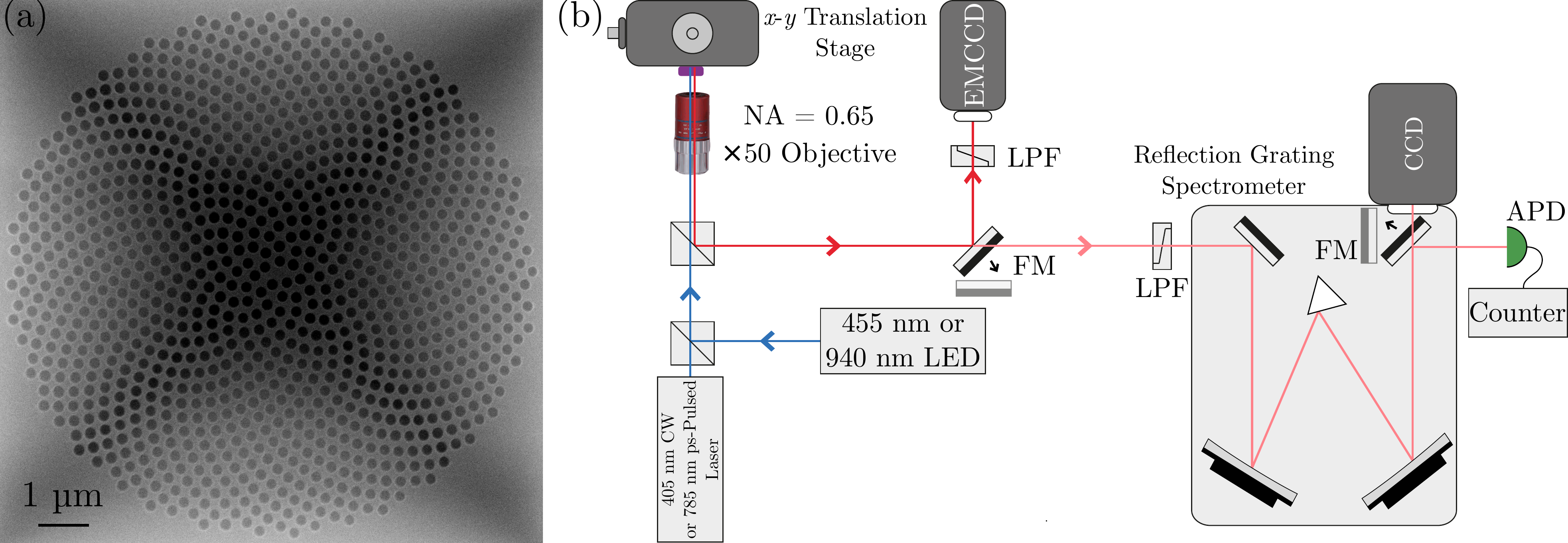}
\caption{Fabricated aperiodic nano-photonic devices and characterisation micro-photoluminescence set-up. (a) Scanning electron micrograph of a suspended silicon nitride aperiodic photonic device with golden-angle Vogel-geometry. 
(b) Schematic of the confocal micro-photoluminescence set-up (not to scale), comprising a light emitting diode (LED) with emission wavelength centred at 455\,nm (and one at 940\,nm for device illumination), a continuous wave (CW) laser emitting at 405\,nm and picosecond(ps)-pulsed 785\,nm laser as excitation sources, focused by a 50$\times$ microscope objective (with numerical aperture NA\,=\,0.65) onto a sample placed on an $xy$-translation stage, within a cryostat for cryogenic measurements. The detection is carried out by an Electron Multiplying Charge Coupled Device (EMCCD) for photo-luminescence imaging, by a CCD for spectral characterisation and by an avalanche photo-diode (APD) with photon counting electronics, at the exit port of a reflection grating spectrometer, for time-resolved measurements. (LPF\,=\,550\,long-pass filter, FM\,=\,flip mirror), the squares represent beam-splitters.}
\label{Fig2}
\end{figure}

\newpage
\begin{figure}[h!]
\centering
\includegraphics[width=\linewidth]{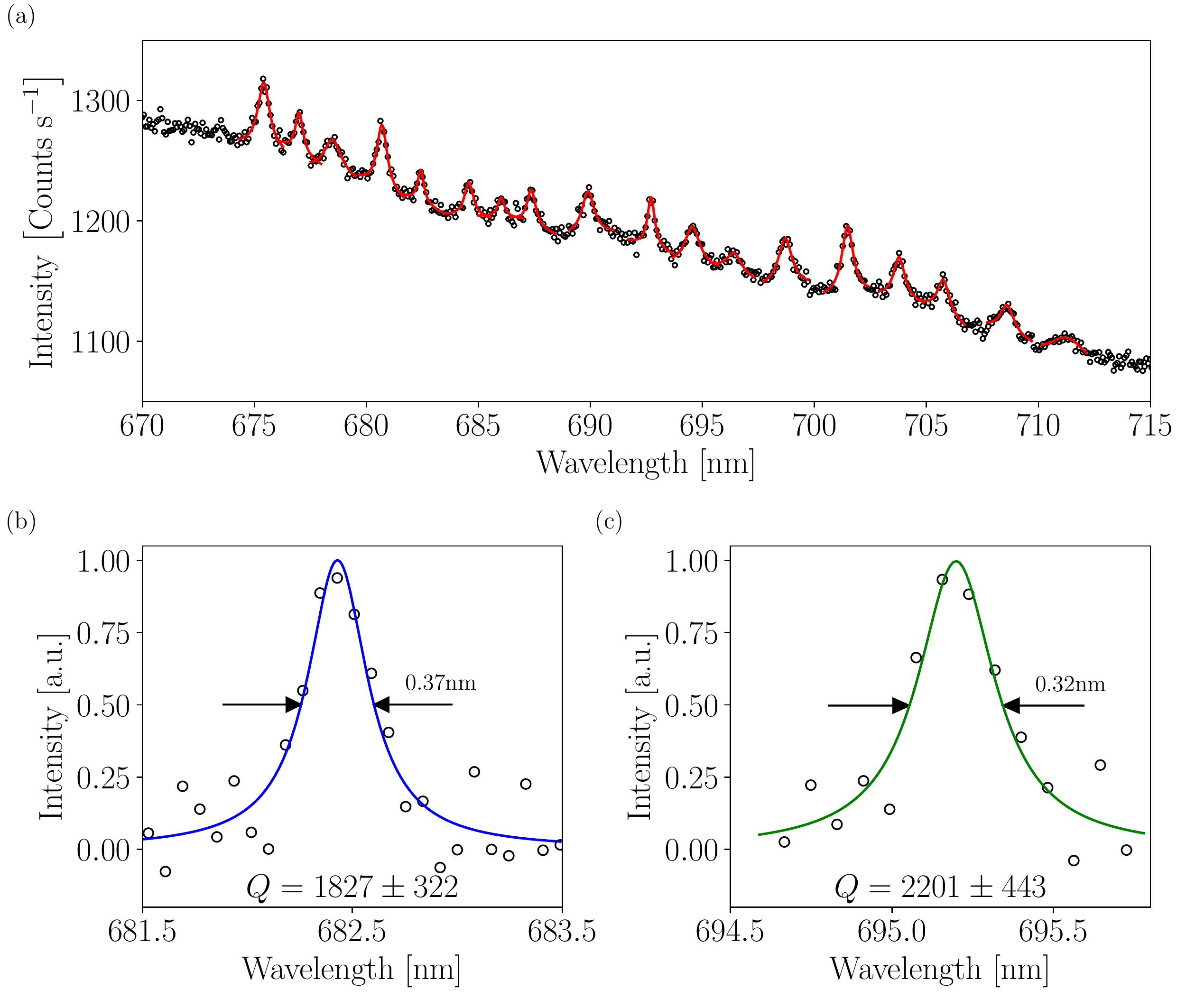}
\caption{Micro-photoluminescence measurements proving efficient light confinement on a Si$_3$N$_4$ chip. (a) Example of a broad-range photo-luminescence spectrum showing sharp optical resonances, signature of light confinement in deterministic aperiodic photonic devices in silicon nitride. The spectrum was collected at room temperature, under 405\,nm CW laser illumination, with a power density of 1.8\,kW/cm$^2$. The solid lines represent Lorentzian fits to the data (symbols). (b,c) Examples of optical resonances, collected under the same conditions as panel (a), showing Lorentzian fits and the extracted quality factors $Q$.}
\label{Fig3}
\end{figure}

\newpage
\begin{figure}[h!]
\centering
\includegraphics[width=\linewidth]{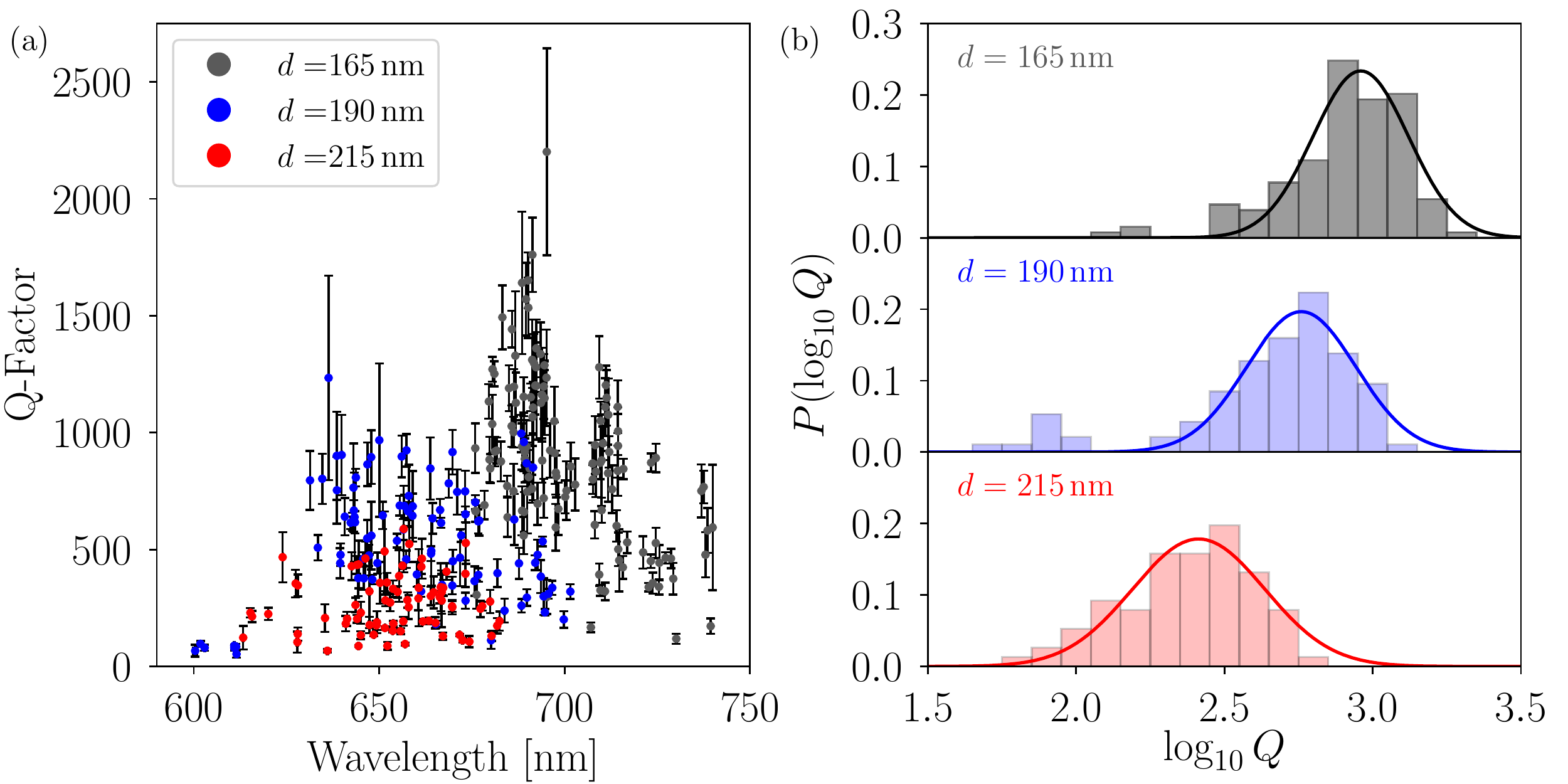}
\caption{Statistical analysis of the optical resonances confined by aperiodic Si$_3$N$_4$ photonic devices. (a) Statistics of the quality factors distributions collected from micro-photoluminescence spectra, like the ones shown in Fig.\,2, plotted as a function of emission wavelength, for aperiodic photonic devices with hole diameter of 165 (black),\,190 (blue),\,215 (red)\,nm. (b) Probability distribution of the quality factors shown in panel (a), plotted with the same colour coding. Coloured continuous lines are log-normal fits to the data.}
\label{Fig4}
\end{figure}

\newpage
\begin{figure}[h!]
\centering
\includegraphics[width=\linewidth]{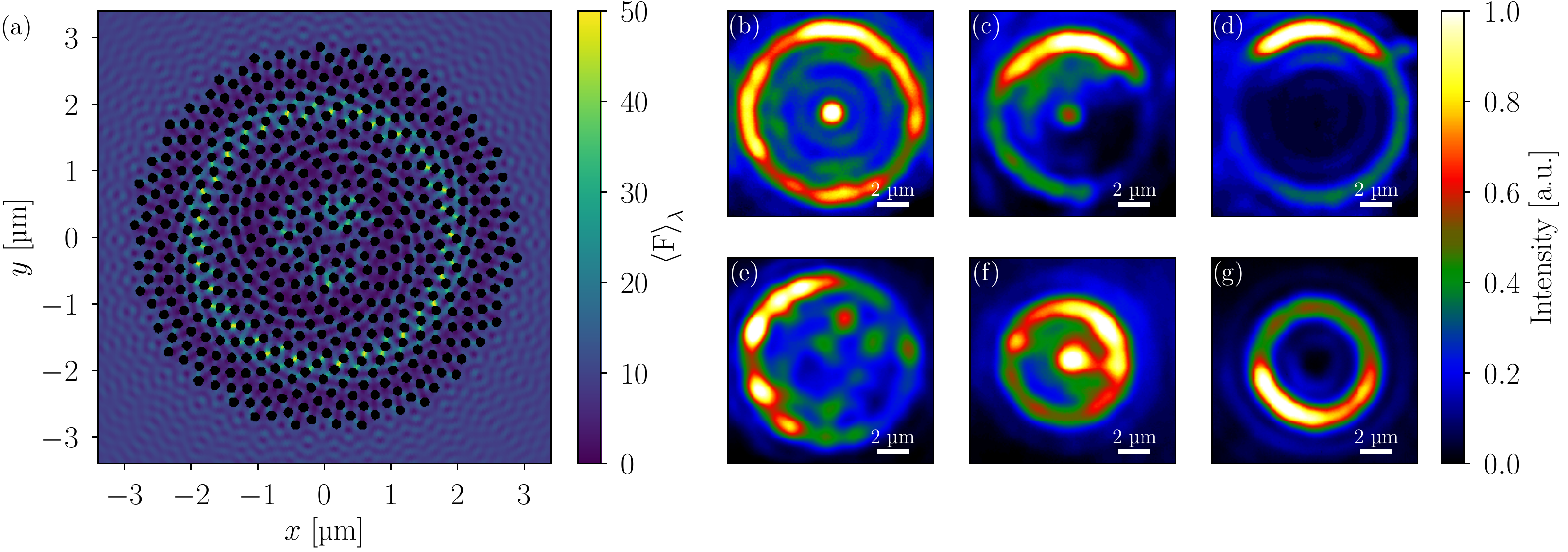}
\caption{Confined optical modes in aperiodic nano-photonic GaAs devices. (a) Simulation of the spectrally integrated LDOS ($\langle\mathrm{F}\rangle_{\lambda}$) superimposed to the aperiodic device structure. (b-g) InAs quantum dots grown in the middle of the GaAs membranes are excited by a 455\,nm LED with a power density of 40\,W/cm$^2$, at a temperature of 10\,K, and the emitted light is collected by an EMCCD, after a 550\,nm long-pass filter. Photo-luminescence images of the optical modes confined by the GaAs aperiodic nano-photonic devices are shown in panel (b) from a device with hole diameter D=190\,nm and average centre-to-centre spacing P=240\,nm, panel (c) from a device with D=160\,nm and P=230\,nm, panel (d) from a device with D=160\,nm and P=240\,nm, panel (e) from a device with D=180\,nm and P=230\,nm, panel (f) from a device with D=180\,nm and P=240\,nm, panel (g) from a device with D=190\,nm and P=230\,nm. }
\label{Fig5}
\end{figure}

\newpage
\begin{figure}[h!]
\centering
\includegraphics[width=\linewidth]{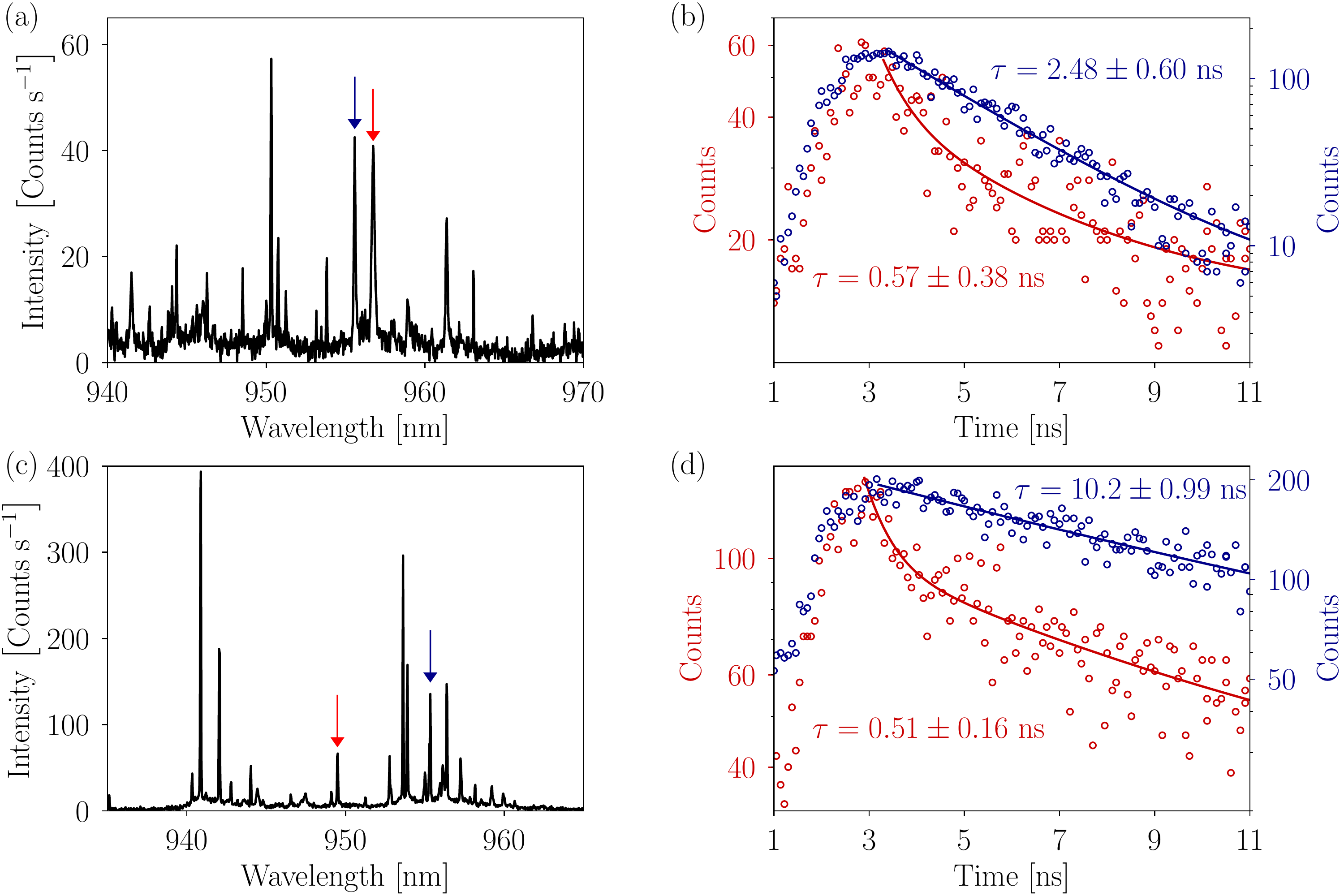}
\caption{Cavity quantum electrodynamics with InAs/GaAs quantum dots embedded within aperiodic nano-photonic devices. (a,c) Photo-luminescence spectra collected on a CCD under 780\,nm laser illumination with power density of 18\,W/cm$^2$ (panels (a,b)) and 339\,W/cm$^2$ (panels (c,d)),  at a temperature of 10\,K. (b,d) Time-resolved measurement of the intensity of the light emitted by the optical transitions highlighted (and colour-coded) by the arrows in panels (a,c), collected by an avalanche photo-diode. Solid lines are exponential fits to the data (symbols) and the extracted decay time characteristics (lifetimes $\tau$) are shown.}
\label{Fig6}
\end{figure}

\newpage

\onecolumngrid \bigskip
\begin{center} {{\bf \large SUPPLEMENTARY INFORMATION}}\end{center}
\setcounter{figure}{0}
\makeatletter
\renewcommand{\thefigure}{S\@arabic\c@figure}
\setcounter{equation}{0}
\makeatletter
\renewcommand{\theequation}{S\@arabic\c@equation}

\section{Two-Dimensional Generalized Lorenz-Mie Theory}
Two-dimensional generalized Lorenz-Mie theory (2DGLMT) is a rigorous spectral method for solving Maxwell's equations in a geometry consisting of an array of non-overlapping, infinite circular cylinders, where waves propagate in the $xy$ plane, orthogonal to the cylinders' axes ($z$ axis by convention). This method has been presented in a number of papers \cite{elsherbeni1992,felbacq1994,nojima2005,andreasen2011}. The presentation here follows the nomenclature used in ref.\cite{gagnon2015} with modifications to allow for interior and exterior line-source (the 2D equivalent of a dipole) according to Ref.\cite{asatryan2003}.

In two-dimensional geometries where propagation is constrained to within the plane, electromagnetic waves can be decomposed into two polarizations: transverse electric (TE) ($\boldsymbol{H}=H_z\hat{\boldsymbol{z}}$) and transverse magnetic (TM) ($\boldsymbol{E}=E_z\hat{\boldsymbol{z}}$). For a photonic device consisting of air holes embedded in a dielectric medium, such as the device that we are considering, a band-gap is expected to occur for TE polarization\cite{joannopoulos2008}. Consequently, the following discussion will be restricted to the case of TE polarization, for which $\boldsymbol{E}=(E_{x}, E_{y},0)$.

As in traditional Mie-theory formulations \cite{bohren1998,jin2015}, the fields are decomposed into exterior and interior components, written as a sum of base functions that reflect the symmetry of the individual scatterers. For cylindrical particles, the basis is given in terms of cylindrical Bessel $J_l(z)$ and Hankel functions of the first kind $H_l^{(+)}(z)$ of order $l$. While the Bessel functions describe the source and interior fields, the Hankel functions express the scattered fields radiating from each cylinder. The total field $ H_{z}^{E}$, consisting of the scattered and incident (exterior) source fields, is given by:

\begin{subequations}
	\begin{align}
    H_{z}^{E}(\bm{r}) &= H_{z}^{sca}(\bm{r}) + H_{z}^{E,inc}(\bm{r})\\[+10pt]
	H_{z}^{E,inc}(\bm{r})  &= \sum_{l=-\infty}^{\infty}a_{nl}^{0E}J_{l}(k_{o}\rho_{n})e^{jl\theta_{n}}  \\[+10pt]
	H_{z}^{sca}(\bm{r}) &= \sum_{n=1}^{N} \sum_{l=-\infty}^{\infty} b_{nl}H_{l}^{(+)}(k_{o}\rho_{n})e^{jl\theta_{n}} 
	\end{align} 
	\label{eq:external}
\end{subequations}
where $H_{z}^{E,inc}$ is the exterior source field (generated by a line-source located in the embedding medium) with base coefficients $a_{nl}^{0E}$, and $H_{z}^{sca}$ is the field scattered by all cylinders with base coefficients $b_{nl}$, known as the Mie-coefficients. Lastly, $(\rho_n,\theta_n)$ are the cylindrical coordinates centered on the $n^{th}$ cylinder. The interior field is likewise written as:

\begin{subequations}
	\begin{align}
    H_{z}^{I}(\bm{r}) &= H_{z}^{int}(\bm{r}) + H_{z}^{I,inc}(\bm{r}) \\[+10pt]
	H_{z}^{I,inc}(\bm{r})  &= \sum_{l=-\infty}^{\infty}a_{nl}^{0I}H_{l}^{(+)}(k_{n}\rho_{n})e^{jl\theta_{n}} \\[+10pt]
	H_{z}^{int}(\bm{r})  &=  \sum_{l=-\infty}^{\infty} c_{nl}J_{l}(k_{o}\rho_{n})e^{jl\theta_{n}}
	\end{align}
	\label{eq:internal}
\end{subequations}
where $H_{z}^{I,inc}$ is the interior source field (generated by a line-source located inside the $n^{th}$ cylinder), while $H_{z}^{int}$ is the interior field with base coefficients $c_{nl}$  describing the transmission (and/or reflection) of fields at the $n^{th}$ cylinder's boundary.

In order to apply boundary conditions to each cylinder, Graf's addition theorem\cite{abramowitz1970} is used to write all fields in the $n^{th}$ cylinders coordinate system. This leads to a convenient matrix equation relating the source and the Mie coefficients:

\begin{equation}
\label{eq:Tmat}
\boldsymbol{\mathrm{T}}(k) \boldsymbol{\mathrm{b}}=\boldsymbol{\mathrm{a}}\\[+10pt]
\end{equation}
where $\boldsymbol{\mathrm{a}}$ is a vector related to $a_{nl}^{0E}$ and $a_{nl}^{0I}$, $\boldsymbol{\mathrm{b}}=b_{nl}$, and $\boldsymbol{\mathrm{T}}(k)$ is the transfer matrix describing the material and geometrical properties of the structure under investigation. The explicit form of $\boldsymbol{\mathrm{T}}(k)$ can be found in Eq. (54) of Ref.\cite{gagnon2015}.

\section{Local density of states and Purcell enhancement}

The local density of optical states (LDOS) quantifies the number of electromagnetic modes into which photons can be emitted at a given position in space. It is particularly useful for comparison to experimental results because it is related to both the total power and the spontaneous decay rate of light emitted by a dipole source as a function of wavelength and position in a heterogeneous dielectric structure\cite{sprik1996,novotny2012principles}. Moreover, comparing the LDOS in a photonic device to the free-space LDOS quantifies how much light emission is enhanced or suppressed due to the presence of the structure.

In 2D, the LDOS is proportional to the total intensity radiated by a line-source\cite{asatryan2003} and it is evaluated from the trace of the two-dimensional electric Green's tensor\cite{balian1971} $\boldsymbol{\mathrm{G}^e}(\boldsymbol{r},\boldsymbol{r'})$:
\begin{equation}\label{eq:LDOS}
\rho(\boldsymbol{r};\lambda)=-\frac{4 n_b^2}{c\lambda}\textrm{Im }  \textrm{Tr}[\boldsymbol{\mathrm{G}}^{e}(\boldsymbol{r},\boldsymbol{r};\lambda)]
\end{equation}
where $n_b$ is the refractive index of the embedding medium. For TE polarization, the trace in Eq.(\ref{eq:LDOS}) reduces to
\begin{equation}
\textrm{Tr}[\boldsymbol{\mathrm{G}}^{e}(\boldsymbol{r},\boldsymbol{r};\lambda)]=\boldsymbol{\mathrm{G}}_{xx}^{e}(\boldsymbol{r},\boldsymbol{r};\lambda)+\boldsymbol{\mathrm{G}}_{yy}^{e}(\boldsymbol{r},\boldsymbol{r};\lambda)
\end{equation}

Calculating the LDOS requires solving Eq.(\ref{eq:Tmat}) with coefficients $\boldsymbol{a}$ defining a line source in a homogeneous medium for both $\boldsymbol{\hat{x}}$ and $\boldsymbol{\hat{y}}$ orientations, and then evaluating and summing the total fields (either using Eqs.(\ref{eq:external}) or (\ref{eq:internal}), depending on where the source is located) at the position of the line source. The LDOS maps for the golden-angle Vogel-spiral membranes discussed in the main text (see Fig.\,1) are calculated by evaluating the LDOS at the center of the structure as the wavelength and hole diameter are varied. 

Another useful quantity is the Purcell enhancement (PE), which describes the modification of the radiative properties of an emitter due to its environment. The Purcell enhancement $\mathrm{F}(\mathbf{r};\lambda)$ can be evaluated by dividing Eq.(\ref{eq:LDOS}) by the LDOS in vacuum in two dimensions $\rho_0(\boldsymbol{r};\lambda)=(c\lambda)^{-1}$:
\begin{equation}\label{PE}
    \mathrm{F}(\mathbf{r};\lambda)=\frac{\rho(\boldsymbol{r};\lambda)}{\rho_0(\boldsymbol{r};\lambda)}=-4n_{b}^{2}\: \mathrm{Im} Tr [\boldsymbol{\mathrm{G}}^{e}(\boldsymbol{r},\boldsymbol{r};\lambda)]
\end{equation}
Since Eq.(\ref{PE}) quantifies the radiated power, it can be used to explain the photo-luminescence image shown in Fig.\,5 of the main text by spatially averaging the Purcell enhancement over the emitting area of the device:
\begin{equation}
    \left< \mathrm{F}\right> _{\Omega}(\lambda)=\frac{1}{A_\Omega}\iint_{\Omega}\rho(\mathbf{r};\lambda)d^{2}\mathbf{r}
\end{equation}
where $\Omega$ identifies the scanned dielectric region of the device with area $A_\Omega$. Note that $\Omega$ excludes the air-holes which do not emit any light. The spectrally integrated emission profiles depicted in Fig.\,5 can similarly be recreated by spectrally averaging  the Purcell enhancement over a bandwidth $\Delta \lambda$:
\begin{equation}
    \left< \mathrm{F}\right> _{\lambda}(\mathbf{r})=\frac{1}{\Delta \lambda}\int_{\bar{\lambda}-\Delta \lambda/2}^{\bar{\lambda}-\Delta \lambda/2}\mathrm{F}(\mathbf{r};\lambda)d\lambda
\end{equation}

\begin{figure}[h!]
	\begin{center}
		\includegraphics[width=\linewidth]{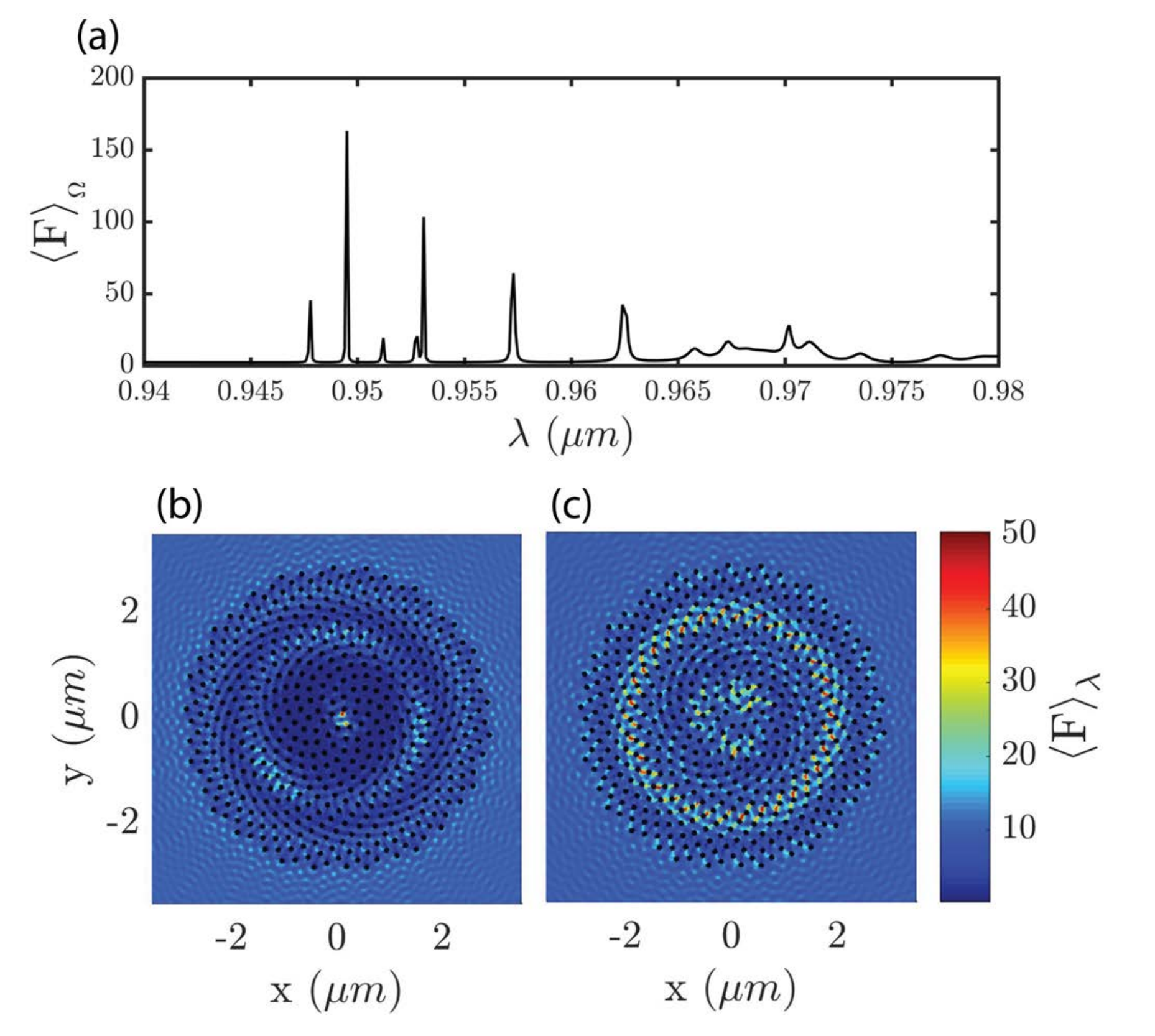}
		\caption{LDOS analysis. (a) Spatially integrated LDOS $\rho_{\Omega}$ as a function of wavelength. (b) and (c) are spectrally integrated LDOS $\rho_{\lambda}$ integrated over 20\,nm bands centered on 944\,nm and 1010\,nm, respectively.}
		\label{fig:purcell}
	\end{center}
\end{figure}

Fig.\ref{fig:purcell} presents the results of this analysis applied to a GaAs membrane with 500 air-holes of 220\,nm diameter arranged in a golden-angle Vogel-spiral geometry characterized by a scaling factor of $a_0=0.13$ $\mu m$ (see eq.(1) in the main text). 
 The LDOS, and subsequently the PE, were evaluated on a spatial grid with 20\,nm spacing and 0.1\,nm spectral resolution where, in comparison to the experiments, each spatial point effectively represents an InAs quantum dot embedded in the membrane. To take into account the out-of-plane losses \cite{Qiu}, we have considered an effective refractive index evaluated from the fundamental guided mode of a 190\,nm GaAs slab surrounded by air. Convergence tests were performed with respect to the multipolar order coefficient $l$, and we have numerically verified that a value of  $l_{max}=4$ is adequate. The field summations run from $-l_{max},...,l_{max}$, thus a total of $2*l_{max}+1=9$ angular orders are considered in the calculation.  The spatially-averaged Purcell enhancement is shown in Fig.\ref{fig:purcell}(a) and consists of sharp oscillating spikes with varying width and height. On the other hand, $\left< \mathrm{F}\right> _{\lambda}$ with a $\Delta \lambda=20$nm integration band centered at $\bar{\lambda}=944$ nm and $\bar{\lambda}=1010$ nm are reported in Fig.\ref{fig:purcell} panel (b) and (c), respectively. These profiles consists of bright annular regions and bright central regions similar to the measurements depicted in Fig.\,5 in the main manuscript. Moreover, both the measured and simulated spatial profiles resemble the spatial map of the first neighbours connectivity obtained from the Delaunay triangulation of this particular spiral \cite{TrevinoOx}. This is a consequence of the unique spatial order of golden-angle Vogel-spirals that defines a radial heterostructure that traps light in regions of different lattice constants, similar to concentric rings of Omniguide Bragg fibers \cite{Steven}.
\section{Mode analysis}
Electromagnetic modes are calculated by solving the source-less Maxwell's equation. In 2DGLMT, this is equivalent to solve the following matrix equation:
\begin{equation}
\label{eq:Tmatmode}
\boldsymbol{\mathrm{T}}(k)\boldsymbol{\mathrm{b}}=\boldsymbol{0}\\[+10pt]
\end{equation}
Solutions of Eq.(\ref{eq:Tmatmode}) only exist when $\textrm{det}[\boldsymbol{\mathrm{T}}(k)])=0$. Therefore, the problem of determining the modes is reduced to the one of  finding the discrete set of complex values $k_m$ for which this condition is met. Subsequently, the spatial field profile is calculated from the eigenvectors $\boldsymbol{\mathrm{b}}$. Solving Eq.(\ref{eq:Tmatmode}) is a nonlinear eigenvalue problem, a subject of ongoing research \cite{Molesky}.

In this work, we have calculated the eigenvalues by generating two-dimensional maps of $\textrm{det}[\boldsymbol{\mathrm{T}}(\lambda,\textrm{Im}(k))]$ with a spectral resolution of $\Delta \lambda$=0.2\,nm. Here $\lambda$ is equal to 2$\pi/\textrm{Re}(k)$. An equal spacing in log-space of 
$\Delta [\textrm{log}_{10} \textrm{Im}(k)]$=0.0156 was considered. The $\boldsymbol{\mathrm{T}}$ matrix was calculated by truncating the multipolar expansion up to $l_{max}$=4 and by considering 500 air-holes with different diameters $D$ equal to 165, 190, 215\,nm arranged in a golden-angle Vogel spiral geometry in silicon nitride. The complex wavenumbers of the modes were then obtained as the position of the local minima of $\det(\boldsymbol{\mathrm{T}})$.

\begin{figure}[h!]
	\begin{center}
		\includegraphics[width=\linewidth]{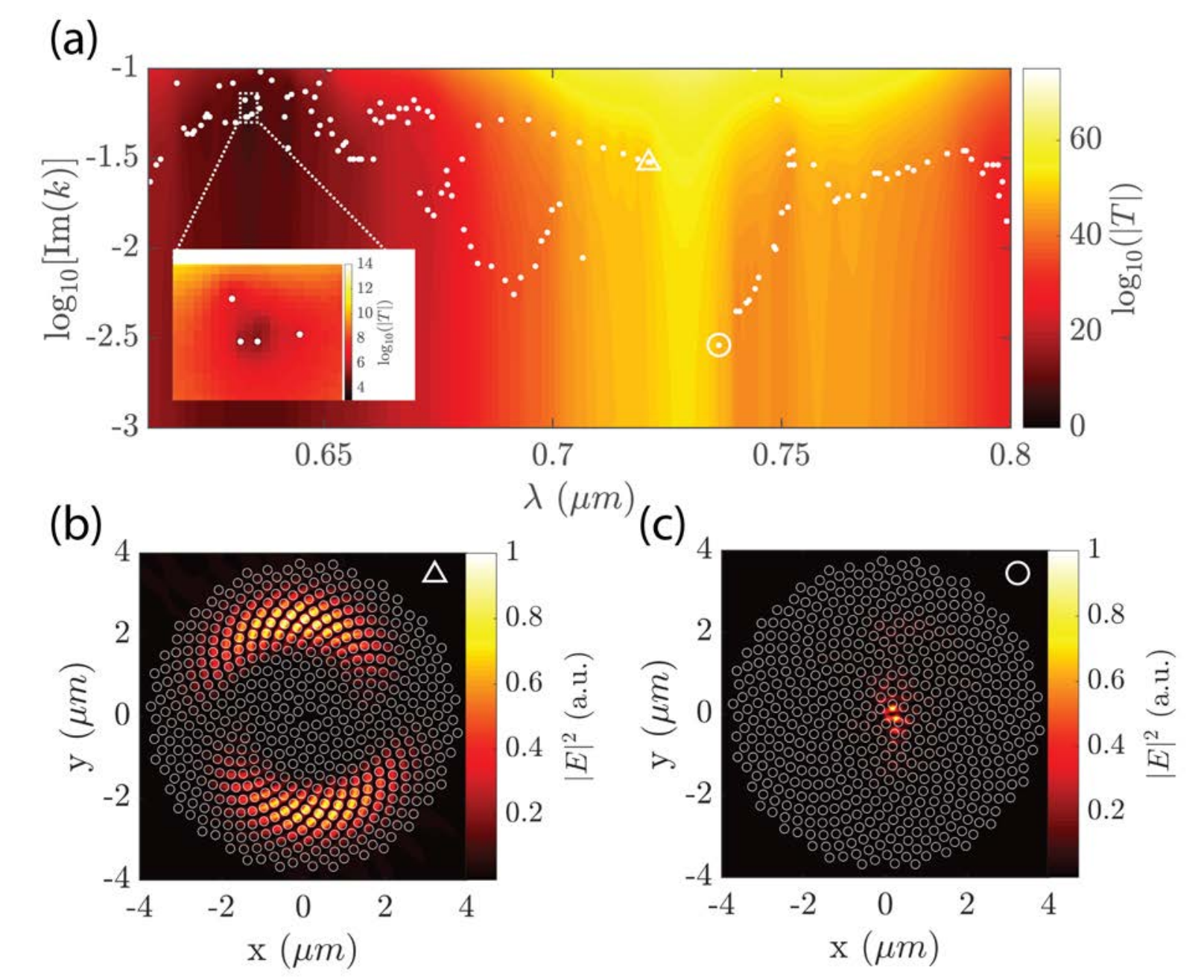}
		\caption{Mode solving example. (a) 2D map of $\textrm{det}[\boldsymbol{T}(\lambda,\textrm{Im}(k))]$, white dots indicate local minima which correspond to complex eigenvalues of the array. Inset shows a zoom-in of an area where the local minima are observable by eye. (b) Electric-field intensity of a left band-edge mode. (c) Electric-field intensity of a right band-edge mode. The complex eigenvalues corresponding to the plotted modes are indicated by the respective shapes in panel (a).}
		\label{fig:Tmap}
	\end{center}
\end{figure}

An example of the T-map obtained in the case $D$=215\,nm is reported in Fig.\ref{fig:Tmap}(a). White dots represent the positions of local minima that are the eigenvalues corresponding to the desired optical modes. The band-gap can be clearly identified by the bright vertical strip around $\lambda \approx 0.73$ $\mu m$ where no modes are present. Fig.\ref{fig:Tmap} (b-c) display the electric field intensity, calculated from the curl of eqns.(\ref{eq:external}) and  (\ref{eq:internal}) of two representative modes located on the left and right band-edge, respectively. As expected, the field of the left band-edge mode is largely concentrated in the air region, while the field of the right band-edge mode is mostly confined in the dielectric medium \cite{joannopoulos2008}. 

\begin{figure}[h!]
	\begin{center}
		\includegraphics[width=\linewidth]{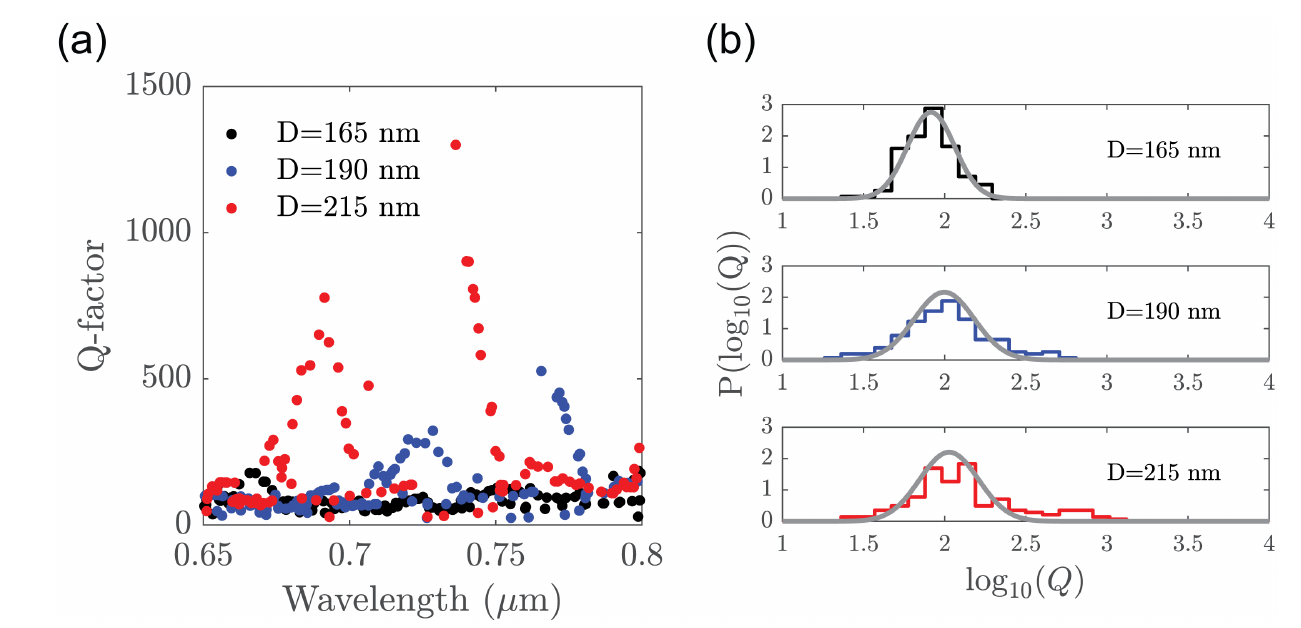}
		\caption{Analysis of the Q factors found using 2DGLMT. (a) Plot of Q vs wavelength as the hole diameter changes. Increasing diameter results in a blue-shift of the band-gap. The band-gap for D=165\,nm appears to the right of the wavelength-region depicted. (b) Log histograms of Q factors shown in panel (a) with log-normal fits.}
		\label{fig:Qplot}
	\end{center}
\end{figure}

The complex wavenumbers $k_m$ of the modes can be used to calculate their quality factors (Q-factors) through the relation\cite{Lalanne,Jackson}:
\begin{equation}
Q=\Bigg|\frac{\textrm{Im}(k)}{2\textrm{Re}(k)}\Bigg|
\end{equation}
The Q-factors of the three device geometries studied in the main text are plotted as a function of wavelength in Fig.\ref{fig:Qplot}(a). The band-gap considerably red-shifts as the air-hole diameter increases. This feature, particularly evident for the $D$=215\, nm configuration, is consistent with the results reported in Fig.1(c-d) of the main text where we show the 1D-LDOS cuts as a function of wavelength. Moreover, Fig.\ref{fig:Qplot}(a) confirms that the high quality factor modes are concentrated near the band-gap \cite{TrevinoOx,Liew,SgringuoliACS,John}. In order to further understand the Q-factor statistics of Fig.\,4, we report in Fig.\ref{fig:Qplot} (b) the histogram of the logarithm of the Q-values. These data are well described by a log-normal probability distribution function (continuous grey-lines). Interestingly, the log-normal distribution of Q-factors has been predicted and observed in uniform media in the regime of Anderson localization \cite{Pinheiro,Weiss,Smolka}. However, differently from traditional random media where the log-normal distribution of Q-factors is associated to exponentially localized modes, optical resonances of Vogel spirals are characterized by a multi-length-scale decaying behavior that is related to the multifractal complexity of their geometrical supports as well as of their LDOS spectra \cite{TrevinoOx}. 

There is a notable difference in the trend of Q-factors with the air-hole size for the measured and simulated data. While simulations predict an increase in Q as hole-diameter increases, the opposite was observed in the measurements (see Fig.\,4 with respect to Fig.\ref{fig:Qplot} (b)). This difference is related to the fact that the measured values are a result of interplay between the in-plane ($Q_{\parallel}$) and out-of-plane ($Q_{\perp}$) contributions to the total Q-factors.  On the one hand, our numerical
simulations can only take into account the in-plane contribution, $Q_{\parallel}$. On the other hand, it is well known that spatially localized modes, such as the one shown in Fig.\ref{fig:Tmap} (c) resembling a defect state of a photonic crystal cavity, have large leakage in the vertical direction \cite{joannopoulos2008,Yamilov,Akahane,boriskina2008}. Therefore, the simulated high-Q modes, that have many $\mathbf{k}$ vectors falling within the leaky region \cite{Akahane}, are not efficiently guided within the membrane but rather they leak out to free-space very quickly \cite{boriskina2008}. This loss channel results in a measured total quality factor that is smaller than the calculated transverse one.

One additional aspect that must be taken into account in the comparison between experiment and simulation is that the fabricated devices consists of 1000 air holes while the highly demanding computational resources limit the numerical simulations to 500 holes. Indeed, the number of holes impacts the number of modes characterizing the spiral \cite{TrevinoOx}. To circumvent this issue, we have studied the scaling properties of the simulated Q-factors for a membrane with 215\,nm holes with respect to the number of spiral elements $N$. Fig.\ref{fig:Qplot} (a) shows the histograms of the logarithm of the Q-values calculated by varying $N$ from 100 (red-curve) up to 500 (cyan-curve) with a step of 100. The histogram of the corresponding measured data is also shown for comparison at the top (black-curve). Log-normal fits (black-continuous lines) are superimposed on these histograms. The mean ($\mu$) and the variance ($\sigma^{2}$) values of the fits are reported above each baseline and confirm a scaling trend. Panel (b) shows the cumulative sums of the histograms as well as the cumulative probability distributions of the log-normal fits. In order to quantify the log-normality of the data, we employed the Shapiro-Wilk (SW) inferential method \cite{shapiro1965}. The SW metrics are all close to unity, indicating the good quality of the log-normal fit. Finally, panel (c) shows plots of all of the simulated Q-factors (black-dots) as a function of $N$ and compares them to the measured (red-dots) values. The blue dots represent the median for each $N$, and a linear fit to the simulated median Q-factors intersects almost perfectly with the median of the measured Q-factors. This extrapolation suggests that simulated Q-factors for an array of 1000 air-holes, which was not computationally feasible, would agree quite well in magnitude with the measurements.

Finally, we assess the scaling properties of localized modes of the structure. To ensure high precision of the mode positions, $\det(\boldsymbol{\mathrm{T}})$ was iteratively calculated with higher resolution about the local minima in Fig.\ref{fig:Tmap} until the associated eigenvalues fell below $10^{-4}$. Fig.\ref{fig:intensityCuts} shows four different mode amplitudes ($|H|$) and radial cuts of their intensity ($|H|^{2}$) profiles. Modes (a) and (b) are representative of localized states in the central region of the structure and manifest power-law scaling with  multiple (e) and single (f) scaling parameters. The localized modes in panels (c) and (d) are radially-confined around R=2 and are characterized by a more complex spatial scaling behavior, as shown in panels (g) and (h), respectively.

%

\newpage

\newpage

\newpage
\begin{figure}[h!]
	\begin{center}
		\includegraphics[width=\linewidth]{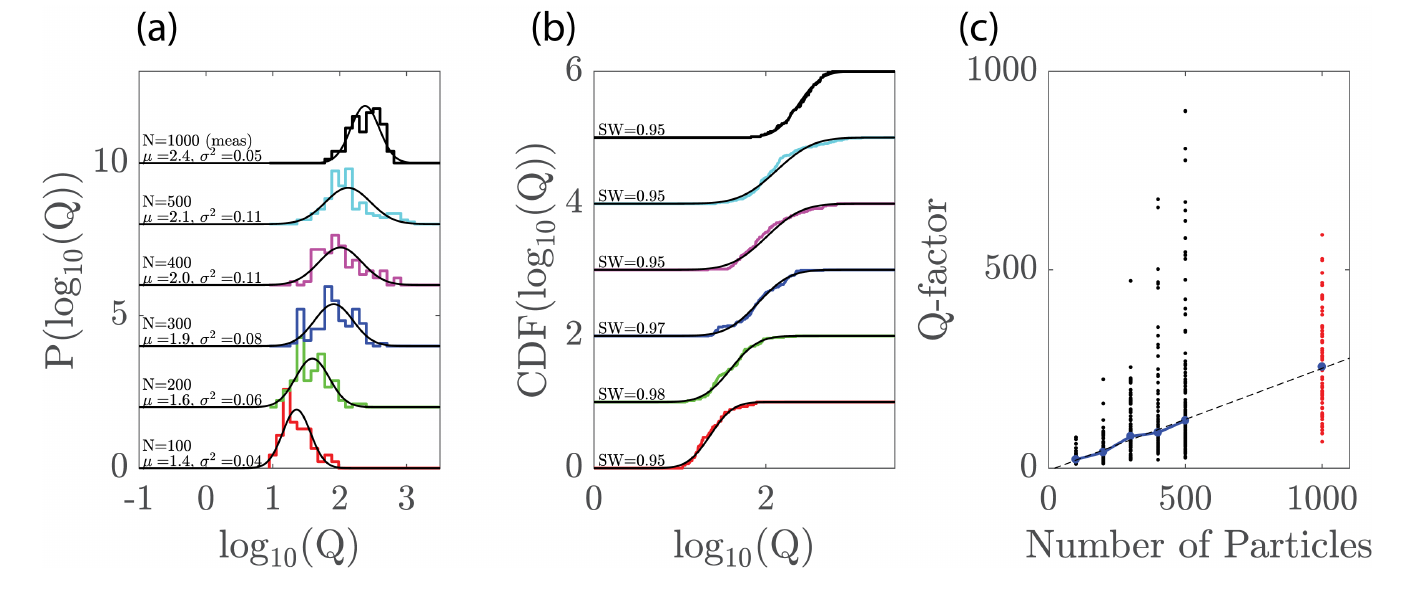}
		\caption{Q factors trend analysis with increasing particle numbers with 215\,nm diameter. (a) $\log_{10}$ histograms of Q-factors for increasing number of particles from 100 to 500. The top (black) data is obtained from experimentally measured Q-factors. Mean and variance of each fit is indicated above the respective baseline. (b) Cumulative $\log_{10}$ histogram with fits. Above each baseline is the Shapiro Wilk (SW) metric, demonstrating the high quality of the fits. (c) Plot of Q-factors against particle number. Black dots are simulated, red dots are measured data, and blue dots are the median of each respective column. The black dashed line represents the trend of the simulated medians.}
		\label{fig:Qtrend}
	\end{center}
\end{figure}

\newpage
\begin{figure}[h!]
\begin{center}
\includegraphics[width=\textwidth]{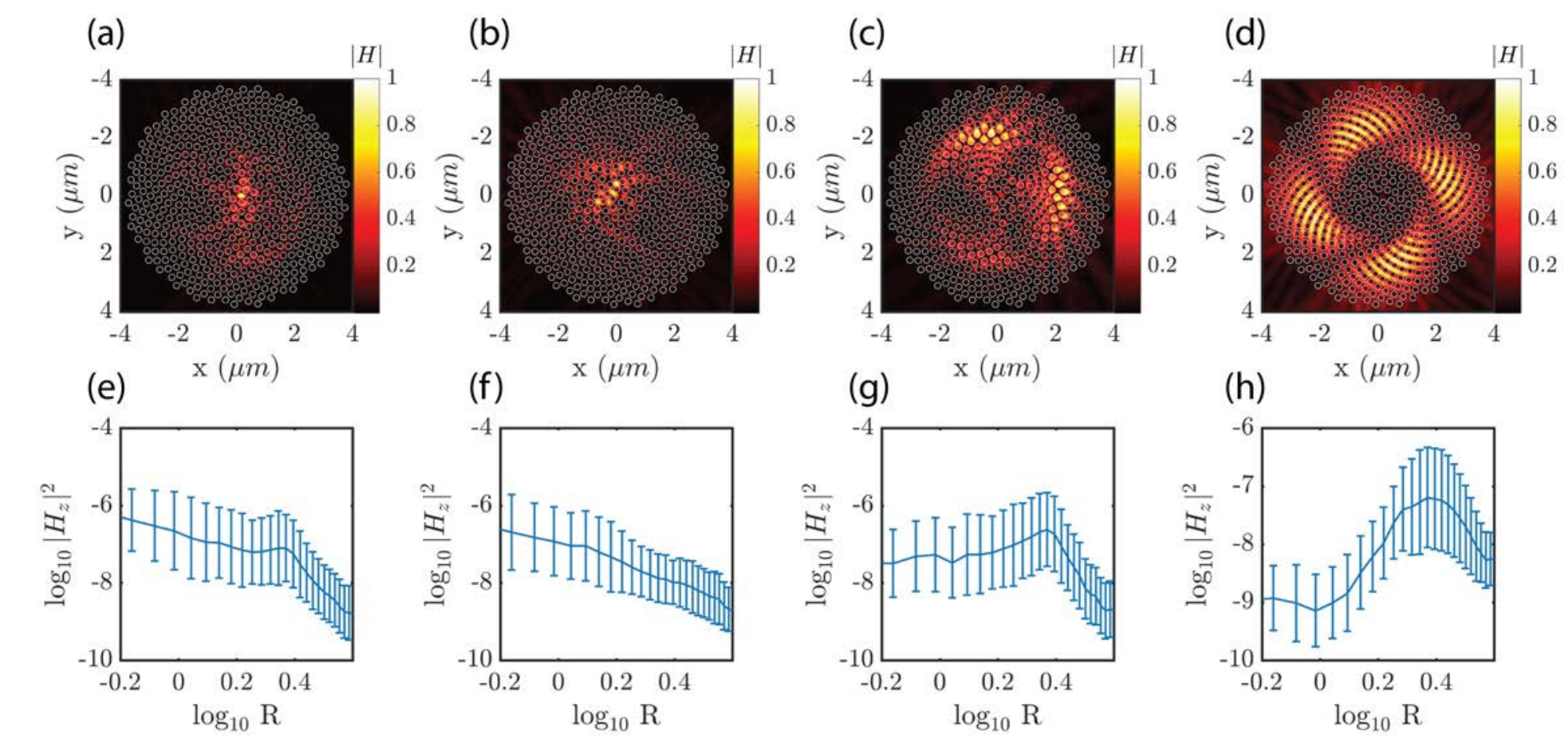}
\caption{(a-d) H-field mode amplitude profiles for a Si$_{3}$N$_{4}$ structure with 500 holes of 215\,nm diameter. The complex wave numbers of the modes are $k=8.0299-0.0031i$, $8.3687-0.0088i$, $7.9817-0.0044i$, and $8.2284-0.0312i$ respectively. All modes are normalized to a peak value of 1. (e-h) Radial cuts of the mean intensity ($|H|^{2}$) as a function of radius $R$ from the center of the spiral, plotted on a log-log scale. Error bars are the standard deviation of $\log_{10}(|H|^2)$.}
\label{fig:intensityCuts}
\end{center}
\end{figure}

\end{document}